\documentclass[epj]{svjour}

\usepackage{graphicx}
\usepackage{amssymb}
\usepackage{amsmath}
\usepackage{fancyhdr}
\def\makeheadbox{{
 }}
\def\mytitle{My title} 
\def\myauthors{My name}  
\def\mytype{My type of session}
\def\mysession{My session}

\def\to{\rightarrow}

\def\bi{\begin{itemize}}
\def\ei{\end{itemize}}
\def\be{\begin{equation}}
\def\ee{\end{equation}}
\def\bea{\begin{eqnarray}}
\def\eea{\end{eqnarray}}

\def\lsim{\mathrel{\rlap{\lower4pt\hbox{\hskip1pt$\sim$}}
    \raise1pt\hbox{$<$}}}                
\def\gsim{\mathrel{\rlap{\lower4pt\hbox{\hskip1pt$\sim$}}
    \raise1pt\hbox{$>$}}}                

\def\mytitle{Collider Phenomenology of Higgsless models} 
\def\myauthors{Alexander Belyaev}    
\def\mytype{Contributed Talk}    
\def\mysession{Alternatives}

\newcommand{\beqa}{\begin{eqnarray}}
\newcommand{\eeqa}{\end{eqnarray}}
\newcommand{\beq}{\begin{equation}}
\newcommand{\eeq}{\end{equation}}
\newcommand{\bq}{\begin{equation}}
\newcommand{\eq}{\end{equation}}
\newcommand{\ba}{\begin{array}}
\newcommand{\ea}{\end{array}}
\def\ga{g_0^{~}}
\def\gaa{g_0^2}
\def\gb{g_1^{~}}
\def\gbb{g_1^2}
\def\gc{g_2^{~}}
\def\gcc{g_2^2}
\def\fa{f_1^{}}
\def\faa{f_1^2}
\def\fb{f_2^{}}
\def\fbb{f_2^2}

\def\f{\frac}
\def\dis{\displaystyle}
\def\[{\left[}
\def\]{\right]}
\def\({\left(}
\def\){\right)}
\def\ga{g_0^{~}}
\def\gaa{g_0^2}
\def\gb{g_1^{~}}
\def\gbb{g_1^2}
\def\gc{g_2^{~}}
\def\gcc{g_2^2}
\def\fa{f_1^{}}
\def\faa{f_1^2}
\def\fb{f_2^{}}
\def\fbb{f_2^2}

\begin{document}
\title{Collider Phenomenology of Higgsless models}
\author{Alexander Belyaev\inst{1}
\thanks{\emph{Email:} a.belyaev@soton.ac.uk}%
\thanks{the study has been done in 
collaboration with 
Hong-Jian He, Yu-Ping Kuang, Yong-Hui Qi, Bin Zhang, R.~Sekhar Chivukula, 
Neil D.~ Christensen, 
Elizabeth H.~Simmons and Alexander Pukhov~\cite{He:2007ge}}
}                     
%
%
\institute{School of Physics and Astronomy, University of Southampton, Southampton, SO17 1BJ, U.K.}

%
\date{}
\abstract{
 We study the LHC signatures of new gauge bosons in the
 minimal deconstruction  Higgsless model (MHLM).
 We analyze the $W'$ signals of $pp\to W' \to WZ$ and
 $pp\to W'jj \to WZjj$ processes at the LHC, including the complete
 signal and background calculation in the gauge invariant
 model and have demonstrated the LHC potential to 
 cover the whole parameter space of the MHLM model.
 }
\maketitle
%

\section{Introduction}
 
 Disentangling the nature of electroweak symmetry breaking (EWSB) 
 is one of the important challenges of  particle physics today
 and upcoming  CERN Large Hadron Collider (LHC), in particular.
 Among several appealing theories of EWSB,
 Higgsless models are especially promissing.
 In particular, those models predict new heavy gauge bosons
 serving as a key for EWSB and delaying unitarity violation
 of longitudinal weak boson scattering~\cite{unitary5D,HeDPF04}
 without invoking a fundamental Higgs scalar.
  Dimensional deconstruction formulation of the Higgsless theories
  is shown to provide their most general 
  {\it gauge-invariant} formulation\,\cite{HeDPF04,DC04a}
  under arbitrary geometry
 of the continuum fifth dimension (5d)
 or its 4d discretization with only a few lattice
 sites\,\cite{4site,3site}.

  The Minimal Higgsless Model (MHLM)  consists of just 3 lattice
 sites (``The Three Site Model'')\,\cite{3site}
 and predicts
 just two extra $W'$ and $Z'$ bosons
 which mass $\gtrsim!400$\,GeV 
 is  consistent with all the precision data\,\cite{3site}.
 MHLM is 
 gauge invariant via spontaneous symmetry breaking and
 predicts just one pair of nearly degenerate new $(W',\,Z')$ bosons,
 unlike any 5d Higgsless models
 with a tower of Kaluza-Klein gauge-states.
 This model contains all the essential ingredients 
 of Higgsless theories being the simplest realistic 
 Higgsless model with distinct collider signatures.
 In this  study we investigate phenomenology of
 MHLM signals at the LHC  including the complete
 signal and background calculation demonstrate
 demonstrated the LHC potential to 
 cover the whole parameter space of the MHLM model.

\section{MHLM model}

 The MHLM\,\cite{3site} is defined as a
 chain moose with 3 lattice sites, under the 
 5-dimensional 
$SU(2)\times SU(2) \times  U(1)$
gauge theory  with electroweak symmetry
breaking encoded in the boundary conditions of the gauge fields.
 Gauge and Goldstone sectors of MHLM have 5 parameters in total --
 3 gauge couplings $(\ga,\,\gb,\,\gc)$ and 2 Goldstone decay constants
 $(\fa,\,\fb)$, satisfying two conditions due to
 its symmetry breaking structure,
 \beqa
 \hspace*{-6mm}                          
 \f{1}{\gaa} + \f{1}{\gbb} + \f{1}{\gcc} ~=~ \f{1}{e^2}\,,
 ~&&~
 \f{1}{\faa} + \f{1}{\fbb} ~=~ \f{1}{v^2}\,.
 \eeqa
 For the optimal delay of unitarity violation
 we choose equal decay constants
 $\,\fa =\fb=\sqrt{2}v$\, where $\,v=\(\sqrt{2}G_F\)^{-1/2}\,$
 as fixed by the Fermi constant.
 Choice of $M_W$ and $M_{W'}$ as inputs 
 allows to determine  $(\ga,\,\gb,\,\gc)$ gauge couplings.
 The MHLM exhibits a delay of unitarity violation
 for  weak boson scattering
 $V_{L}^aV_{L}^b\to V_{L}^cV_{L}^d$ ($V=W,Z$)
 and for $M_{W'}\lesssim 1$\,TeV,
 each elastic $V_{L}V_{L}$ scattering remains
 unitary over the main energy range of the LHC.

 The fermion sector contains SM-like chiral fermions: left-handed
 doublets $\psi_{0L}^{}$ under $SU(2)_0$ and right-handed weak
 singlets $\psi_{2R}^{}$. For each flavor of $\psi_{0L}^{}$, there is
 a heavy vector-fermion doublet $\Psi_1^{}$ under $SU(2)_1$.
 The mass matrix for $\{\psi,\Psi\}$ is\,\cite{3site}
 \beq                                   
 \dis
 M_F = \left(\ba{cc} m & 0 \\[2mm]
                     M & m' \ea\right)
 \equiv M\left(\ba{cc} \epsilon_L^{} & 0 \\[2mm]
                                   1 & \epsilon_R^{}  \ea\right) \,.
 \eeq
 The mass-diagonalization of $M_{F}$ yields
 a nearly massless SM-like light fermion $F_0$
 and a heavy new fermion $F_1$ of mass
 $\,M_{F_1} = M\sqrt{1+\epsilon_L^2}\,$.
 The light SM fermions acquire small masses proportional to
 $\epsilon_R^{}$\,.\,
 For the present high energy scattering analysis
 we only need to consider light SM fermions
 relevant to the proton structure functions at the LHC, which
 can be treated as massless to good accuracy. So we will set
 \,$\epsilon_R^{}\simeq 0$,\, implying that
 $\psi_{2R}$ and $\Psi_{1R}$ do not mix.
 
 One should stress that fermion sector plays a crucial role in the MHLM.
 First of all, fermion gauge couplings
 in the MHLM\,\cite{3site} are the key to ensure an exact
 gauge-invariance in our collider study contrary 
 to previous studies\,\cite{mat}.
 Secondly,  the fermion sector is the key 
 which provide consistency of MHLM with precision electroweak data.
 The proper adjustment of amount of delocalization of 
 fermions to amount of delocalization of gauge bosons 
 fixes  $\epsilon_L^{}$   via the ideal
 fermion delocalization\,\cite{ideal} (IDDL) condition
 and
 leads to vanishing $W'$-SM fermion couplings and
 thus zero electroweak precision corrections
 at tree-level\,\cite{3site,ideal}.
 
 One should notice that the mass of the heavy
 fermions is strongly  bounded from below to be
 $\,M_{F_1} > 1.8$\,TeV \cite{3site}.
 Therefore the essential phenomenology of MHLM at the LHC
 is related to signals form new gauge bosons $W'$ and $Z'$
 which can be as light as $\sim\!400$\,GeV.
 To simplify the analysis we  consistently
 decouple the heavy fermion by taking the limit
 $\,(M,\,m)\to\infty\,$ while keeping the ratio
 $\,\epsilon_L^{}\equiv m/M$ finite.
 This finite ratio $\epsilon_L^{}$ will be fixed via IDDL~\cite{ideal}.
 
 We have implemented MHLM model
 into CalcHEP package~\cite{calchep} using LanHEP program ~\cite{lanhep} for 
 automatic Feynman rules generation. This implementation has been consistently
 cross-checked in t'Hooft Feynman and Unitary gauges and publicly available
 at\\
 \verb|http://hep.pa.msu.edu/belyaev/public/3-site/|.

 \section{Phenomenology of MHLM}
 
 As discussed above, in MHLM the couplings of new heavy bosons to SM
 fermions are highly suppressed to satisfy precision EW data
 while  the couplings of new heavy bosons to SM gauge bosons are 
 non-vanishing to provide the delay of unitarity for
 $V_{L}^aV_{L}^b\to V_{L}^cV_{L}^d$ amplitudes.
 
 These two essential features
 define the phenomenology of not only MHLM  but the whole 
 class of the Higgsless extradimensional models (HLEDM)
 whose phenomenology will be dominated by the first KK-mode.
 
 In MHLM, the decay width of $W'$ or $Z'$
 are defined by their  decays to $WZ$ 
 or $WW$ pairs, respectively
  \beqa                                      
 \label{eq:W1-width1}
 \Gamma_{V'\to WW(WZ)}
 &=& \f{\alpha M_{V'}}
   {48 s_Z^2 x^2}\[1+O(x^2)\]
 \eeqa
  where $\,\alpha =e^2/4\pi$\, and \,$x\equiv 2M_W/M_{W'}$.\,
 For $M_{W'}=(0.5-1)$\,TeV one has $\,\Gamma_{W'}\simeq (5-31)\,$\,GeV.
  On the other hand, under the IDDL
 $W'$ does not decay to light SM fermions while $Z'$ decay to SM-fermions is
 highly suppressed
  \beqa                                      
 \label{eq:Z1-width1}
 \Gamma_{Z'\to e^+e^-}
 &=& \f{5\alpha M_{V'}x^2 s_Z^2}
   {96 c_Z^4}\[1+O(x^2)\].
 \eeqa
Therefore, one can expect, that the most promising 
discovery channels would be $Z'(W')$
production via gauge couplings with SM gauge bosons.
Moreover, $W'$ production looks more favourable  since
the minimal number of neutrinos 
after  $W'$ leptonic decay is one ($W'\to WZ\to 3l\nu$),
while  $Z'\to WW\to 2l2\nu$ decay channel ends with two neutrinos
disabling the  reconstruction of the $Z'$ peak.

\begin{figure}[h]
 \vspace*{-5mm}
 \includegraphics[width=0.45\textwidth]{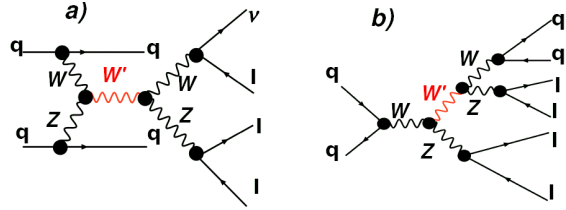}
 \caption{Representative diagrams
  for  the associated $W'Z$
($pp \to W'Z \to WZZ\to 4\ell 2q $) production (a)
 as well as  $W'$
production in $WZ\to W$ fusion process 
($pp \to W'qq \to WZqq\to 3\ell\nu 2q $) (b).
 }
   \label{fig:signal}
 \end{figure}

\begin{figure}[h]
 \vspace*{-5mm}
 \includegraphics[width=0.45\textwidth]{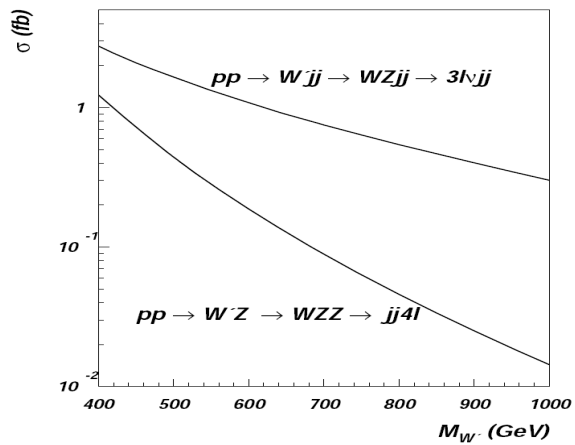}
 \caption{
 The cross sections versus $M_{W'}$ for $pp \to W'Z$ and $pp\to W'qq^{(')}j$
processes including  $4\ell 2q $ and $3\ell\nu 2q$ respective branching ratios.
For $pp\to
W'qq^{(')}$ process $E_q  > 300$~GeV,  $P_Tq>30$~GeV and
$|\Delta\eta_{jj}|>4$ cuts were applied.}
 \vspace*{-0.5cm}
  \label{fig:cs}
 \end{figure}

We found that the most favorable signal processes
for discovery of the MHLM at the LHC  are the associated $W'Z$
($pp \to W'Z \to WZZ\to 4\ell 2q $) production
as well as  $W'$
production in $WZ\to W$ fusion process 
($pp \to W'qq \to WZqq\to 3\ell\nu 2q $)
representative Feynman diagrams
for which are shown in Fig.~\ref{fig:signal}a) and Fig.~\ref{fig:signal}b),
respectively.
The cross sections versus $M_{W'}$ for $pp \to W'Z$ and $pp\to W'qq^{(')}j$
processes including $4\ell 2q $ and $3\ell\nu 2q$ respective branching ratios
are presented in Fig.~\ref{fig:cs}. For $pp\to
W'qq^{(')}j$ process the quark energy ($E_q  > 300$~GeV),  $P_Tq$($P_Tq>30$~GeV) and
rapidity gap cuts were applied $|\Delta\eta_{jj}|>4$. These cuts
are essential for the background suppression as we discuss below. Hereafter
we use CTEQ6L~\cite{cteq6} parton density function and QCD scale  $Q=\sqrt{\hat{s}} $
and $Q=M_Z$ for   $pp\to W'qq^{(')}$ and   $pp\to W'qq^{(')}j$ processes,
respectively.

 As we have mentioned above,
 we propose to analyze the
 $pp \to W'Z \to WZZ$ process via
 leptonic decays of the two $Z$ bosons and hadronic decays of $W$
 providing a clean  signature of 4-leptons plus 2-jets,
 $jj4\ell$\, ($\ell=e,\mu$).
 The backgrounds include:
 (a) the irreducible SM production of
 $pp\to WZZ\to jj4\ell$,
 (b) the reducible background of the SM production,
 $\,pp\to ZZZ\to jj4\ell$,
 with one $Z\to jj$ (mis-identified as $W$
 due to finite experimental di-jet mass resolution)
 and
 (c) the SM process $pp\to jj4\ell$
 other than (a) and (b), which
 also includes the $jj 4\ell$ backgrounds with $jj=qg,gg$.

\begin{figure}[h]
 \vspace*{-3mm}
 \includegraphics[width=0.45\textwidth]{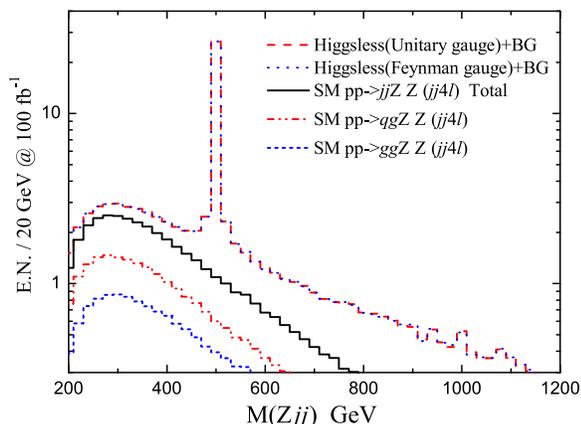}
 \vspace*{-2mm}
 \caption{Signal and background events in the process
 $pp\to W'^{(\ast)}Z \to WZZ\to jj\,\ell^+\ell^-\ell^+\ell^-$
 for an integrated luminosity  of 100\,fb$^{-1}$.}
  \label{M(WZ)2}
\vspace*{-3mm}
\end{figure}

 To suppress  backgrounds we impose the cuts,
 \begin{eqnarray}                            
 && M_{jj} \,=\, 80\pm 15\,{\rm GeV},
 ~~~~~~\Delta R(jj)\,<\,1.5\,,
 \nonumber\\
 && \sum_Z p_T^{~}(Z)+\sum_j p_T^{~}(j) \,=\, \pm 15~{\rm GeV}.
 \label{supp-cuts}
 \end{eqnarray}
 The first cut selects di-jets arising from on-shell $W$ decay to be
 within the experimental resolution\,\cite{ATLAS}; the second cut requires the
 dijet separation of the signal; and the third cut uses the
 conservation of transverse momentum in the signal to suppress the
 background.
 Furthermore, we impose the following electron and jet
 ID/acceptance cuts
 \begin{eqnarray}                          
 p_{T\ell}^{~}>10\,{\rm GeV},~~&&~~ |\eta_\ell^{~}| <
 2.5\,,\nonumber\\
  p_{Tj}^{~}>15\,{\rm GeV},~~&&~~ |\eta_j^{~}| < 4.5\,.
\label{det-cuts}
 \end{eqnarray}
 In Fig.\,\ref{M(WZ)2} we present 
 the $M_{Zjj}$ event distributions for the signal and background
 under these cuts for an integrated luminosity of 100\,fb$^{-1}$. 
 We depict the signal by a dashed curve,
 the backgrounds (c) with $jj=gg,qg$ by dashed and
 dashed-double-dotted curves, respectively,
 and the total background (a)$+$(b)$+$(c)
 by a solid curve.
 The backgrounds (a) and (b) are so small that they are
 not visible in Fig.\,\ref{M(WZ)2}.
  Finally we have chosen 
  $M_{Zjj}=M_{W'}\pm 0.04M_{W'}$
  mass window cut to estimate signal significance and LHC reach.
  In this mass window we have summed 
  contributions from two $Z$ bosons
  for signal and background.
 The gauge-invariance of this calculation is verified by comparing
 the signal distributions in unitary and 't\,Hooft-Feynman gauges;
 as shown in Fig.\,\ref{M(WZ)2} by red-dashed and blue-dotted curves,
 they perfectly coincide.
 %
  From the calculated number of  signal and background events,
 we derive the statistical significance
 from the Poisson probability in the conventional way.
 The integrated luminosity required for detecting the $W'$
 in this channel will be summarized in Fig.~\ref{IL}.

 Next, we analyze the LHC potential to discover  $W'$-boson in the
 $pp \to WZqq'$ process, where the signal is
 given by the $W'$ contribution to
 $WZ\to WZ$ scattering subprocess.
 We perform a complete analysis of
 \,$pp\to WZjj$, and choose the pure leptonic decay modes of
 $WZ$ with 3 leptons plus missing-$E_T$ \cite{bagger,wwh}.
 We carry out a full tree-level calculation including both
 signal and background together.

 To effectively  suppress  $qq \to WZ$
 and  $pp \to WZ jj$  ($jj = qg,\,gg$) QCD backgrounds
 we apply the following
 jets rapidity gap cut and large jet energy cut
 \begin{eqnarray}
 \left|\Delta\eta_{jj}^{}\right| > 4\,,  &&
  E_j^{}    > 300\,{\rm GeV}\,
 \label{newcuts}
 \end{eqnarray}
 in addition to acceptance cuts given by
 \begin{eqnarray}
  p_{Tj}^{} >  30\,{\rm GeV}\,,        && |\eta_j^{}| < 4.5\,\nonumber\\
  p_{T\ell}^{~} \,>\, 10\,{\rm GeV}\,,  && |\eta_{\ell}^{~}| \,<\, 2.5
\label{jj rap diff}
 \label{oldcuts}
 \end{eqnarray}
 where $E_j^{}$ and $p_{Tj(l)}^{}$ are transverse energy and
 momentum of  final-state jet(lepton),
 $\eta_{j(l)}^{}$ is the jet(lepton) rapidity, and
 $\left|\Delta\eta_{jj}^{}\right|$ is
 the difference between the rapidities of the two forward jets.
  For computing the SM EW backgrounds, we need to
 specify the reference value of the SM Higgs
 mass $M_H$. Because the SM Higgs scalar only contributes to the
 $t$-channel in $pp\to qq'WZ $, we find that varying the Higgs
 mass in its full range $M_H=115\,{\rm GeV}-1\,$TeV has little
 effect on the SM background curve.
 Hence we can simply set $M_H=115$\,GeV in our
 plots without losing generality.

  Using  cuts (\ref{newcuts})-(\ref{oldcuts})
  we have computed the $W Z$
 invariant mass ($M_{WZ}$) distribution in both unitary gauge and
 't\,Hooft-Feynman gauge  and have revealed 
 an extremely precise and large cancellation
 between the fusion and non-fusion contributions for $pp \to W Z qq'$ process, 
 as required by the exact gauge-invariance.
  These cancellations cannot be inferred
 without a truly gauge-invariant model, contrary to the approach of
 imposing only a naive 5d sum rule\,\cite{mat}.
 Traditional analyses\,\cite{bagger} of gauge-boson fusion in a
 strongly-interacting symmetry breaking sector have relied on using
 separate calculations of the signal and background
 while in our case in the correct gauge-invariant implementation of MHLM
 we can perform direct calculation of $qq \to WZ q q'$ 
 process for  signal and background together.
 \begin{figure}[h]
  \includegraphics[width=0.45\textwidth]{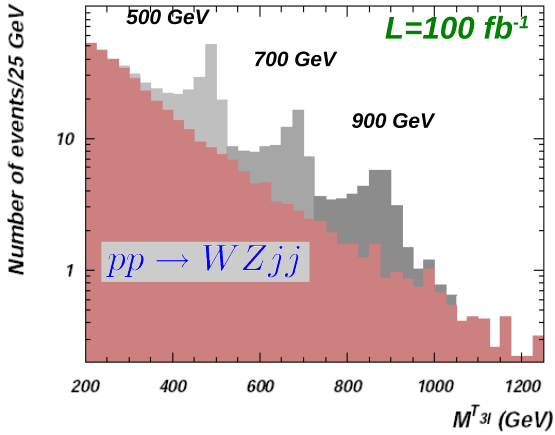}
 \caption{Numbers of signal and background events versus
 the transverse mass $M_T(WZ)$ after imposing the cuts
 (\ref{newcuts})-(\ref{oldcuts})
 for an integrated luminosity of 100\,fb$^{-1}$.}
  \vspace*{-3mm}
 \label{fig:wzjj}
 \end{figure}
 
 Since there is just one neutrino in $3\ell\nu$ signature,
 we can use transverse mass variable,
 $M_T^2(WZ)=$\\
 $[\sqrt{M^2(\ell\ell\ell)+p_T^2(\ell\ell\ell)}
    +|p_T^{\rm miss}|]^2-|p_T^{}(\ell\ell\ell)+p_T^{\;\rm miss}|^2$ \cite{bagger}
 for the effective signal over the background rejection.
 In Fig.~\ref{fig:wzjj} we present $M_T^2(WZ)$ distributions
 for  $M_{Z'}=500,700,900$~GeV exhibiting clear Jacobian peaks.
 We compute signal significance for the $0.85M_{W'}<M_T<1.05M_{W'}$
 window and obtain the required integrated luminosities
 for $3\sigma$ and $5\sigma$ detections of the $W'$ boson 
 presented
 in Fig.\,\ref{IL}.
\begin{figure}[h]
 \includegraphics[width=0.45\textwidth]{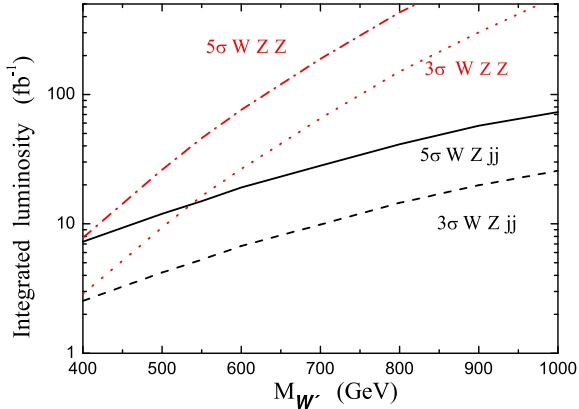}
 \vspace*{-0.2cm}
 \caption{Integrated luminosities required
 for $3\sigma$ and $5\sigma$ detection
 of $W'$ signals as a function of $M_{W'}$.
 The dotted and dashed-dotted curves are for the $WZZ$
 channel, while the dashed and solid curves are
 for the $WZjj$ channel.}
 \label{IL}
\vspace*{-0.4cm}
  \end{figure}

\section{Summary}
 We present the first study on  LHC potential 
 to observe signatures predicted by 
 Minimal Higgsless Model (MHLM)\,\cite{3site}.
 We have calculated the complete 
 gauge invariant signals and backgrounds for two
 promising processes, $pp \to WZZ\to jj\,4\ell$ and
 $pp \to WZjj \to \nu 3\ell\,jj$.
 In this analysis, we only take account of the statistical error.
 We have checked that typical electromagnetic ($0.1/\sqrt{E(GeV)}$) 
 and hadronic($0.5/\sqrt{E(GeV)}$) 
 detector energy resolution~\cite{ATLAS} which we have approximated by 
 gaussian  smearing, has a very small
 effect on the presented distributions (Fig.3,4) and final results (Fig.5).
 Other issues related to the details of detectors, such as the
 systematic error, detection efficiency,
 etc, are beyond this  study.
 Both $WZZ$ and  $WZqq'$ channels have clean leptons signatures
 and reconstructable $W'$ mass. With the proposed cuts we can 
 effectively suppress all SM backgrounds.
 We would like to stress that the calculation in the context of an exactly
 gauge-invariant Higgsless model (such as the MHLM\,\cite{3site}),
 is vital for analyzing the $pp \to WZjj$ process consistently.
 One should also stress the complementarity of $WZZ$ and  $WZqq'$  channels.
 The first one provide clean resonance peak and would allow the precise 
 reconstruction of the $W'$ mass
 while  the second one has larger cross section
 and with 100 fb$^{-1}$ integrated luminosity
 would allow to  completely cover 
 MHLM parameter space up to unitarity limit at $M_{W'}\simeq 1.2$~TeV. 
 We summarize the
 $3\sigma$ and $5\sigma$ detection potential of the LHC
 in Fig.\,\ref{IL} where the required integrated luminosities are shown.
 For example,  for $M_{W'}=500\,(400)$\,GeV, the $5\sigma$
 discovery of $W'$ requires an integrated luminosity
 of 26\,(7.8)\,fb$^{-1}$
 for $pp \to WZZ\to jj\,4\ell$, and
 12\,(7)\,fb$^{-1}$ for $pp \to  WZ jj \to \nu3\ell\,jj$.
 These are within the reach of the first few years run at the LHC.
 The evidence for both signals from the $W'$ boson,
 as well as the {\it absence} of a Higgs-like signal in
 $pp\to ZZqq\to 4\ell \,qq$, will be strong evidence for
 Higgsless electroweak symmetry breaking.
 
 To conclude, for the first time we have consistently studied 
 MHLM model which is very well motivated and 
 has several appealing features:
 it is simple but generic,
 since the phenomenology of any Higgsless extradimensional model 
 is  dominated by 
 the first KK-mode;
 the perturbatively calculable MHLM could shed a light on  
 its conjectured dual strongly interacting theory;
 MHLM consistently implements the first KK-mode
 in a gauge-invariant way;
 MHLM satisfies precision EW measurements,
 suggests a very distinctive phenomenology while
 its parameter space, as we have shown, is fully testable at the LHC.
\\
 {\bf Acknowledgments:}
A.~B. thanks SUSY 2007 organizers  for warm hospitality.

\end{document}